%% file: access.tex
\documentclass[10pt, twocolumn]{IEEEtran}
\IEEEoverridecommandlockouts
\usepackage{cite}
\usepackage{amsmath,amssymb,amsfonts}
\usepackage{algorithmic}
\usepackage{graphicx}
\usepackage{textcomp}
\usepackage{color}

\usepackage[
singlelinecheck=false % <-- important
]{caption}
\usepackage{cite}
\usepackage{amsmath,amssymb,amsfonts}
\usepackage{algorithmic}
\usepackage{graphicx}
\usepackage{textcomp}
\usepackage{subfigure}
\usepackage{paralist}
\usepackage{color}
\usepackage{url}
\usepackage{amsmath}
\usepackage{amssymb}
\usepackage{graphicx}
\usepackage{bm}
\usepackage{paralist}
\usepackage{subfigure}
\usepackage{times}
\usepackage{balance}
\usepackage{cite}
\usepackage{comment}
\usepackage[font=small,labelfont=bf,tableposition=top]{caption}
\usepackage{epsfig}

\usepackage[ruled,boxed,vlined]{algorithm2e}

\usepackage{array}
\newcolumntype{P}[1]{>{\centering\arraybackslash}p{#1}}
\newcolumntype{M}[1]{>{\centering\arraybackslash}m{#1}}

\input{mymath}
\usepackage{makecell}

\newcommand\etal{{\em et al.}}
\newcommand{\bnote}[1]{{\color{black} #1}}

\renewcommand{\ie
}{{\em i.e., }}

\renewcommand{\etal}{{\em et. al.\mbox{$\:$}}}

\renewcommand{\baselinestretch}{0.99}

\def\BibTeX{{\rm B\kern-.05em{\sc i\kern-.025em b}\kern-.08em
    T\kern-.1667em\lower.7ex\hbox{E}\kern-.125emX}}

\usepackage{multirow}
\usepackage{subfigure}

\newcommand{\nc}{\newcommand}

\newcommand{\ua}{\uparrow}

\usepackage[utf8]{inputenc}
\usepackage{fourier} 
\usepackage{array}
\usepackage{makecell}

\def\cA{{\cal A}}

\def\cS{{\cal S}}
\def\BibTeX{{\rm B\kern-.05em{\sc i\kern-.025em b}\kern-.08em
    T\kern-.1667em\lower.7ex\hbox{E}\kern-.125emX}}

\nc{\da}{\downarrow} 
\nc{\hc}{\hat{c}} 
\nc{\hS}{\hat{S}}
\nc{\bra}{\langle} 
\nc{\ket}{\rangle} 
\nc{\eq}{equation (\ref}
\nc{\h}{\hat} 
\nc{\hT}{\h{T}}
\nc{\rd}{\textrm{d}}
\nc{\e}{eqnarray}
\nc{\hR}{\hat{R}}
\nc{\Tr}{\mathrm{Tr}}
\nc{\tS}{\tilde{S}}
\nc{\tr}{\mathrm{tr}}
\nc{\8}{\infty}
\nc{\lgs}{\bra\ua,\phi|}
\nc{\rgs}{|\ua,\phi\ket}
\nc{\hU}{\hat{U}}
\nc{\lfs}{\bra\phi|}
\nc{\rfs}{|\phi\ket}
\nc{\hZ}{\hat{Z}}
\nc{\hd}{\hat{d}}
\nc{\mD}{\mathcal{D}}
\nc{\bd}{\bar{d}}
\nc{\mc}{\mathcal}
\nc{\ea}{eqnarray}
\nc{\mG}{\mathcal{G}}

\begin{document}

\title{Twitter Attribute Classification with Q-Learning on Bitcoin Price Prediction
\thanks{$^*$ Corresponding author. School of Computing, Gachon University, Republic of Korea. (email: jychoi19@gachon.ac.kr) }}
              
\author{Sattarov Otabek and 
Jaeyoung Choi$^*$ 
}

\maketitle

\begin{abstract}
\bnote{Bitcoin price prediction based on people's opinions on Twitter usually requires millions of tweets, using different text mining techniques, and developing a machine learning model to perform the prediction. These attempts lead to the employment of a significant amount of computer power, central processing unit (CPU) utilization, random-access memory (RAM) usage, and time. To address this issue, in this paper, we consider a classification of tweet attributes that effects on price changes and computer resource usage levels while obtaining an accurate price prediction. To classify tweet attributes having a high effect on price movement, we collect all Bitcoin-related tweets posted in a certain period and divide them into four categories based on the following tweet attributes: $(i)$ the number of followers of the tweet poster, $(ii)$ the number of comments on the tweet, $(iii)$ the number of likes, and $(iv)$ the number of retweets. We separately train and test by using the Q-learning model with the above four categorized sets of tweets and find the best accurate prediction among them. We compare our approach with a classic approach where all Bitcoin-related tweets are used as input data for the model, by analyzing the CPU workloads, RAM usage, memory, time, and prediction accuracy. The results show that tweets posted by users with the most followers have the most influence on a future price, and their utilization leads to spending 80\% less time, 88.8\% less CPU consumption, and 12.5\% more accurate predictions compared with the classic approach.}

\end{abstract}

\section{Introduction}
\label{sec:introduction}

Earlier stock market forecasting research relied on past stock values \cite{Sattarov2020a,Mallikarjuna19,Atsalakis2009}. Most studies have discovered that analyzing previous prices is not sufficient to anticipate stock market changes because stock market prices are highly volatile. According to the efficient market hypothesis \cite{Fama1965}, financial market movements are influenced by news, current events, and product releases, all of which have a substantial impact on a company's stock value.
As large stock market, Bitcoin has no central controlling authority and is regulated solely by the public. As a result, Bitcoin is viewed as a volatile cryptocurrency and its value is influencing by public ideas. According to the analysis of Kristoufek \cite{Kristoufek2015}, several significant reductions have occured in the Bitcoin exchange rate and in its price during dramatic events in China. Another study conducted by the American Institute for Economic Research \cite{Bitcoina} shows that Bitcoin prices fluctuated substantially between 2016 and 2017 as a result of global news and emotions.

Owing to the rise of social media, information regarding popular sentiments has become more accessible. Social media is becoming an ideal medium for sharing public mood on any issue, and it has a significant effect on general public opinion. Twitter, a social networking service (SNS), has recently received significant academic attention. Twitter is a real-time micro-blogging service that allows users to follow and comment on others' thoughts and views \cite{Leskovec2007}. Approximately 140 million tweets are sent to more than a million people daily. Each tweet is 140-characters long and expresses the public view of a particular issue. Information derived from tweets is valuable for forecasting \cite{Pak2010}.
Over a million Bitcoin-related tweets are available to researchers for processing and application in the field of predicting future Bitcoin prices. %Several sentiment analysis tools are currently in use for detecting opinions within textual data. 
Processing a large amount of Bitcoin-related tweets normally consumes a high level of computer resources (CPU, RAM, memory) and time \cite{Steinkraus2015,Catanzaro2018,McNally2018,Sumarsih2018}. Most of the previous works is focused on how to reduce the resource, so maximizing the prediction result at the same time is not considered.
\begin{figure}[t!]
\begin{center} \centering
\includegraphics[width=1\linewidth]{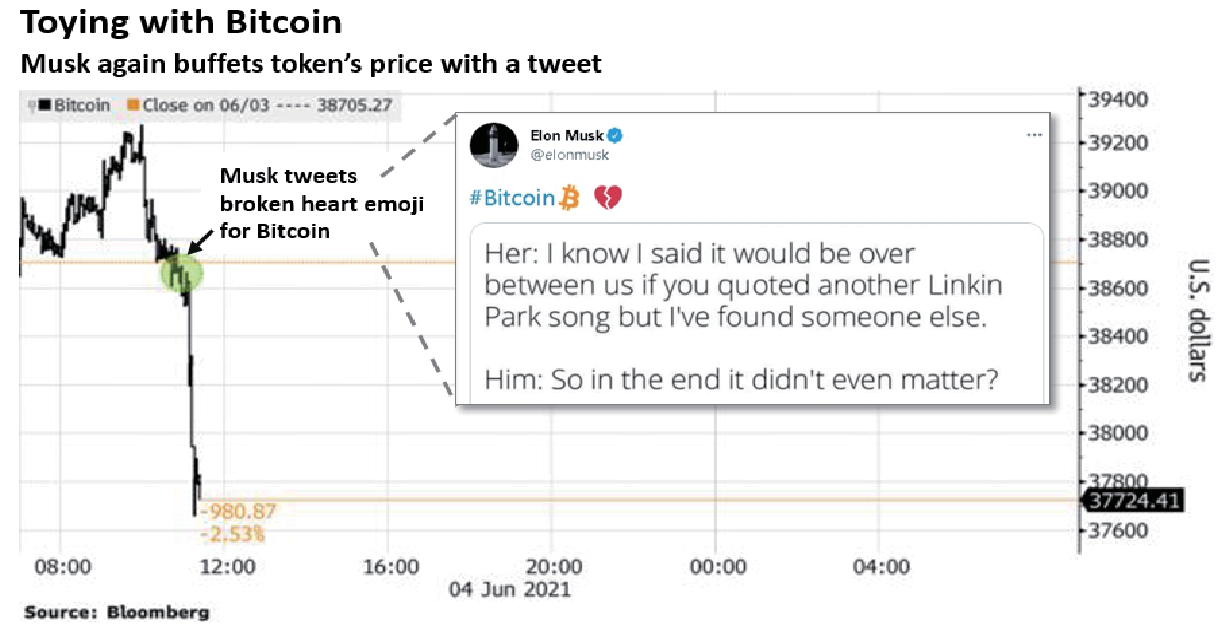}
\caption{Example of Elon Musk's (Tesla CEO) tweets affecting the Bitcoin price \cite{Twitter0}. }
\label{fig:Highway}
\vspace{-0.2cm}
\end{center}
\end{figure}
% \bnote{One of the way to define more effective tweets is focus on how many times tweet is being discussed by other users.}
However, tweets written by an expert, public figure, or celebrity will become viral, with many replies, likes, and retweets. Tweets with few replies, likes, or retweets are unlikely to become viral because they are likely to circulate mainly among close friends. Consequently, viral tweets are expected to have a greater influence on price changes than other tweets. If we can separate tweets with the highest impact on future price changes from less important tweets, it gives the possibility to employ less computer resources usage while still obtaining accurate forecasts.

Hence, different from the previous approaches, in this study, we analyze how Bitcoin-related information on Twitter affects the actual Bitcoin price by considering four main attributes: $(i)$  \noindent{\bf the number of followers of the poster,} $(ii)$  \noindent{\bf the number of comments on a tweet,} $(iii)$  \noindent{\bf the number of likes,} and $(iv)$  \noindent{\bf the number of retweets.}
%Following that, we compare our approach to the traditional approach in which all Bitcoin-related tweets are utilized without being attribute-filtered, by looking at the CPU workload, RAM utilization, memory usage, required time to finish the same task, and forecast accuracy. 
For this, we gather all Bitcoin-related tweets within a particular period and divide them into four groups based on their attributes. 
Since we use the sentiment information of tweets as a resource for the prediction, yet there was no particular guidance to inform the model in what condition of sentiments the price will increase or decrease. Therefore, we need an optimal policy to achieve valuable prediction accuracy. The model can improve its initial non-optimal policy by receiving good/bad rewards based on the prediction results. Considering the above, we choose to use a well-known Q-learning method to obtain the most valuable attribute to predict Bitcoin's future price. To the best of our knowledge, our work is the first study on the predictable range of tweet attributes involving the term ``Bitcoin'' on the future returns and volatility of Bitcoin. 

After classifying which attribute is helpful to separate highly effective tweets to make a prediction, we compare our approach to the classic approaches in which all Bitcoin-related tweets are utilized without being attribute-filtered, by looking at the CPU workload, RAM utilization, memory usage, required time to complete the same task, and prediction accuracy.

We summarize our main contributions in more detail as follows: 
\begin{itemize}
\item[$(a)$] First, we study the predictive power of four main tweet attributes: number of tweet poster's followers, number of comments, number of likes, and number of retweets. We create four datasets consisting of tweets sorted according to the above attributes. Next, we extract the sentiment of each tweet. By making four separate predictions based on the datasets and evaluating the prediction results, we detect the most useful attribute for the Bitcoin price prediction.         

\smallskip
\item[$(b)$] Second, we develop the predictive model based on the Q-learning algorithm. For this, we first consider a Markov Decision Process (MDP) as follows: the current actual price of Bitcoin serves as a state, the prediction of Bitcoin price as an action, and the difference between the actual price and predicted price as a reward. In general, the state transition probability is often not provided which leads for us to adopt the model-free version of Reinforcement Learning (RL). \bnote{Using this, we design several reward functions to improve the prediction accuracy of the Q-leaning.}    

\smallskip
\item[$(c)$]  Finally, we improve the accuracy of prediction and minimize computer resources (CPU, RAM, and memory) utilization and researcher time. The Q-learning based model receives two different datasets as input data where the first dataset consists of all Bitcoin-related tweets without being attribute-filtered (classic approach) and the second dataset is the most useful dataset which we determined earlier among four datasets (proposed approach). With two different datasets, the model gives two different prediction outputs. By comparing the predictions' parameters, we get a conclusion about which approach is better one.         

% \smallskip
% \item[$(d)$]  Finally, we extract effective tweets for the price prediction among all Bitcoin-related tweets. The aforesaid attribute-filtering method gives an opportunity to classify Bitcoin-related tweets into the most useful and less useful categories. To keep the computer from wasting its resources and make an accurate price prediction within a short time, it is better to focus on the most useful tweets rather than paying attention to every single tweet.
\end{itemize}

\smallskip
The remainder of this paper is organized as follows. Section II provides an overview of related research. Detailed information about the data collection, data preprocessing, and sentiment analysis, are provided in Section III. Section IV describes the model learning algorithm and its employment in our research. The experimental results are detailed in Section V. Section VI summarizes the limitations of the study and points the direction for further research. Finally, Section VII concludes the paper.

%\bnote{The section was modified with information that considers the usage of PC resources.}
\begingroup
\setlength{\tabcolsep}{5pt} % Default value: 6pt
\renewcommand{\arraystretch}{2.0} % Default value: 1
\begin{table*}
\caption{Taxonomy of related works}
\centering
\label{taxonomy}
\begin{tabular}{c|l} 
\hline
\textbf{Goal}& \multicolumn{1}{c}{\textbf{Methods}} \\ 
\hline
\begin{tabular}[c]{p{0.20\textwidth}}Bitcoin price prediction with public opinion \end{tabular}          & \begin{tabular}[c]{p{0.60\textwidth}}Pearson correlation \cite{Kaminski2016}, Bitcoin/Etherium \cite{Ranasinghe2021}, NewsSentiment \cite{Nagar2012}, Word2vec and \qquad N-gram \cite{Pagolu2016}, Cumulative sentiment \cite{Kee2017}, Random Forest Regression \cite{Sattarov2020b},  \noindent{\bf Q-Learning [This paper].}\end{tabular}\\ 
\hline
\begin{tabular}[c]{p{0.20\textwidth}}Striving for accurate prediction\end{tabular}             & \begin{tabular}[c]{p{0.60\textwidth}}Linear Regression, RNN, and LSTM \cite{Mittal2019}, \cite{Pant2018}, \cite{Ye2022},\cite{Thanekar2019}, ARIMAX \cite{Serafini2020}, Linear discriminant analysis \cite{Gurrib2021}, ARIMA \cite{Raju2020}, Logistic regression, Naive Bayes, SVM \cite{Colianni2015}, Multiple linear regression \cite{Jain2018}, Tweet corpus in COVID-19 era \cite{Pano2020}, Random Forest, Decision tree, AdaBoost \cite{Luo2020}, XGBoost-Composite model \cite{Ibrahim2021},  \noindent{\bf Q-Learning [This paper].}\end{tabular}  \\  
\hline
\begin{tabular}[c]{p{0.20\textwidth}}Resource usage minimization\end{tabular}     & \begin{tabular}[c]{p{0.60\textwidth}}GPU over CPU \cite{Steinkraus2015}, \cite{McNally2018}, GPU based system for SVM \cite{Catanzaro2018}, Apache Spark \cite{Sumarsih2018},  \noindent{\bf Q-Learning [This paper].}\end{tabular}\\
\hline
\end{tabular}
\end{table*}
\endgroup

\section{RELATED WORK}
In this section, we classify the related researches into the following three categories: (1) Bitcoin price prediction with public opinion, (2) Striving for accurate prediction, and (3) Resource usage minimization.
Table \ref{taxonomy} shows general information about related studies along with the key algorithms/methods they used.

\subsection{Bitcoin price prediction with public opinion}
Sentiment analysis is an important field for researchers, as people's thoughts and emotions have become popular and an acceptable technique for examining and analyzing public opinion. Twitter, Facebook, and Instagram are examples of social media platform used to collect sentiment data for research. The major goal of adopting these approaches is to identify and extract emotions in spoken or written language using natural language processing techniques. Among other social media platforms, Twitter has recently attracted interest from a wide range of academic disciplines, as it is considered useful for analyzing economic and social datasets. Employment of machine learning algorithms on the data extracted from Twitter has opened widely opportunities including identification of hatred speeches \cite{Zia2016}, analyzing personalities based on profile pictures \cite{Bhatti2017}, prediction on offensiveness in tweets \cite{Imran2017}, etc.  

Over the past decade, there have been some studies within the field of finding the links between price movements and sentiments extracted from Twitter. Kaminski \etal \cite{Kaminski2016} found that the platform appears to have an impact on users and information dissemination. Ranasinghe \etal \cite{Ranasinghe2021} demonstrated that Twitter may be related to a shift in the public image of Bitcoin. According to this research, there is a strong link between the probability of Twitter users' influence and the probability of being influenced, but the majority of users maintain a balance in terms of their attitudes in both circumstances. Nagar \etal \cite{Nagar2012} claimed that the sentiment of news obtained from the news corpus and stock price movements were highly correlated. Pagolu \etal \cite{Pagolu2016} focused on forecasting stock price movements using Twitter sentiment, and revealed a strong connection between sentiments on Twitter and stock market movements. Sul \etal \cite{Kee2017} developed a sentiment classifier and compared it with stock returns in 2.5 million tweets related to S\&P 500 companies. The findings revealed that rapid sentiment was more likely to be reflected in a stock price on the same trading day, whereas slower-spreading sentiment was more likely to be reflected on upcoming trading days. In our previous research \cite{Sattarov2020b}, we scrapped more than 9.2 thousand tweets that were posted in a two-month period, and found that when sentiment analysis was applied to tweets regarding Bitcoin and financial data, the sentiment on Twitter had a predictive impact on the Bitcoin findings.

\subsection{Striving for accurate prediction}
It is known that tweet sentiments have positive relationships with price fluctuations. Based on this fact,  several techniques have been proposed to accurately predict the future price by the employment of different machine learning algorithms. Mittal \etal \cite{Mittal2019} gathered approximately 7.5 million tweets and obtained results on tweet sentiment after applying long short-term memory (LSTM), recurrent neural network (RNN), and Polynomial regression, whereas tweet volume and Google trends predicted accuracy of 77.01 percent and 66.66 percent for the Bitcoin direction, respectively. 
Pant \etal \cite{Pant2018} conducted an another RNN model which categorized Bitcoin tweets as good/positive or negative. They used the percentage of them coupled with historical price of Bitcoin. The results showed total 77.62 percent of prediction accuracy. 

While many studies that investigated the token economics based on the Bitcoin network, several researches was focused to analyze the network sentiment on the overall price of Bitcoin. Serafini \etal \cite{Serafini2020} compared two models used for Bitcoin time-series predictions: the Auto-Regressive Integrated Moving Average with eXogenous input (ARIMAX) and RNN. The flow of studies that adopted LSTM to make a price prediction has been continued by Ye \etal \cite{Ye2022}. As an ensemble method along with LSTM, they used gate recurrent unit (GRU). The results showed that their model performance achieved 88.74\% value based on real data from September 2017 to January 2021. 
 
Thanekar \etal \cite{Thanekar2019} demonstrated that artificial intellegence (AI) models using sentiment analysis of tweets containing the keywords ``bitcoin'' or ``btc'' predicted the volatility in Bitcoin values with higher accuracy than models that compared the values without sentiment analysis using machine learning through an autoregressive integrated moving average model and LSTM network. Gurrib \etal \cite{Gurrib2021} achieved 0.828 accuracy in forecasting the next-day price direction by using linear discriminant analysis (LDA) with sentiment analysis of Bitcoin-related tweets. Another study \cite{Raju2020} compared AutoRegressive Integrated Moving Average (ARIMA) and LSTM model to make a real-time prediction of Bitcoin price using public sentiments in tweets and achieved more accurate results by using LSTM. Colianni \etal \cite{Colianni2015} studied how tweet sentiments may be utilized to influence investment decisions, focusing on Bitcoin. The authors employed supervised machine learning algorithms to achieve an hour-by-hour and day-by-day accuracy of above 90\%. Similar with above researchers, Jain \etal \cite{Jain2018} focused on current tweets by classifying positive, negative, and neutral sentiments and accumulating their numbers every two hour to predict the price of Bitcoin and Litecoin two hours in advance. Using multiple linear regression (MLR) model, they utilized more than 1.8 million Bitcoin-related and Litecoin-related tweets to investigate whether social factors were capable of predicting the future price of cryptocurrencies. The study notes that MLR model predicts the price of the Bitcoin and Litecoin with the score of 44\% and 59\% respectively.

As Bitcoin has no central authority to control and its fluctuations are relevant to ongoing news and events, some researchers have studied how COVID-19 outbreak data (number of new cases, recovery, and deaths) can impact the future price of Bitcoin. Pano \etal \cite{Pano2020} provided a corpus of tweet text for Bitcoin-related tweets during the summer of the COVID-19 period. This dataset is publicly available and considers three months to perform unimpeded research. In order to make an accurate price prediction, Luo \etal \cite{Luo2020} tried to feed four different machine learning models with three different data: Bitcoin exchange data, COVID-19 data, and Twitter data from January 2020 to July 2020. One of the findings of this study is COVID-19 data does not help to improve the prediction.

\subsection{Resource usage minimization}
Many researchers have studied how to minimize PC-resource employment while keeping the same working accuracy. One of such study, by Steinkraus \etal \cite{Steinkraus2015} reported over three times faster training and testing processes when the model was implemented on a graphic processing unit (GPU) rather than a CPU. A greater comparison difference was reported by Catanzaro \etal \cite{Catanzaro2018} where the classification time and speed were eight times faster when implementing support vector machine (SVM) on a GPU than when implementing an alternative SVM algorithm that ran on a CPU. In contrast to the above two studies, McNally \etal \cite{McNally2018} ran LSTM model on a CPU and GPU to ascertain the accuracy of the direction of the Bitcoin price in USD. They reported the GPU outperforming by a result of 67.7\%. As the dataset for the model to learn increases, Sumarsih \etal \cite{Sumarsih2018} compared GPU performance with the Apache Spark cluster, which is an in-memory data processing engine that uses RAM instead of an I/O disk. Their data processing simulation using linear regression (LR) to learn Bitcoin trading showed faster results when run on the Apache Spark cluster.  

The common point of all the aforementioned researches is that they considered all types of tweets related to cryptocurrency, without considering the importance of the tweet attributes on price movements. 
To the best of our knowledge, our work is the first attempt to classify the tweet attributes involving the term "Bitcoin" and "BTC", that have effects on the future volatility of Bitcoin price.

%\subsection{Data Preprocessing}
\begin{figure*}[]
\begin{center} \centering
\includegraphics[width=1\linewidth]{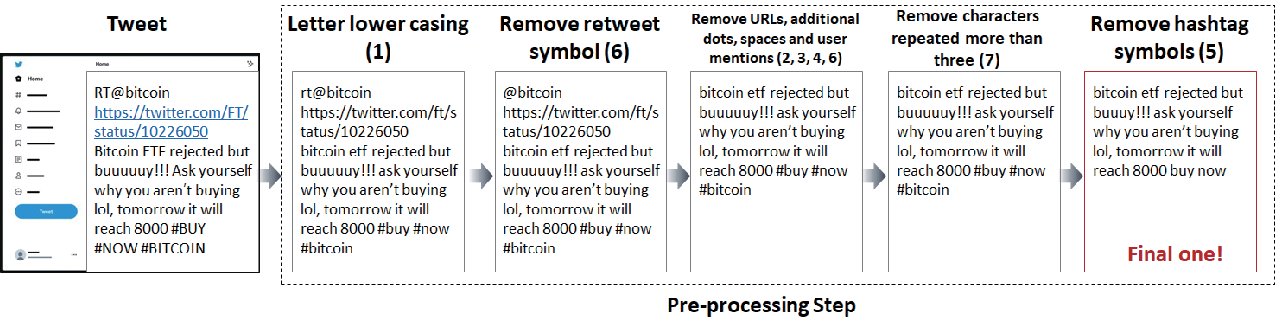}
\caption{Example of data preprocessing tasks in our approach. (At each task, the number in parentheses represents the number in Table~\ref{preprocessing}.) }
\label{fig:preprocessing}
\vspace{-0.1cm}
\end{center}
\end{figure*}
\section{Data Preparation}
\label{sec:guidelines}
In this section, we describe the data-preparation steps for Bitcoin price prediction. We consider the following four steps in data preparation: $(i)$ \noindent{\bf data collection}, $(ii)$ \noindent{\bf preprocessing}, $(iii)$ \noindent{\bf attribute division}, and $(iv)$ \noindent{\bf sentiment analysis}. In the data collection step, we collect data containing tweets relating to Bitcoin. Thereafter, we remove noise such as repeated tweets, URLs, user mentions, and extra repeated characters from the dataset in the preprocessing step. In the attribute division step, we build four datasets containing tweets sorted according to their attributes. We perform sentiment analysis on the gathered tweets in the final sentiment analysis step. The detailed explanation of each step is provided below. 

\subsection{Data Collection}
\noindent{\bf \em Bitcoin Price Data.} We use a total of 1690 days' data that is in the time period from April 1, 2014 to November 14, 2018, in the Bitcoin price market (see \cite{Bitcoinb}) as real data to predict the Bitcoin price because it was observed that the Bitcoin price fluctuated substantially during this period. This motivates us to verify the effectiveness of the proposed method during this period. 

\smallskip
\noindent{\bf \em Bitcoin Tweet Data.} We use Tweepy and Twitter's streaming API \cite{Twitter} for the Bitcoin-related tweet data. Tweepy is a Python-based open-source framework, makes it easier to gather tweets using Twitter API \cite{Tweepy}. Tweepy allows data filtering based on hashtags or terms, which is an effective means of collecting relevant data. The filter keywords are selected using the most definitive Bitcoin context phrases; for example, ``cryptocurrency'' may contain attitudes towards other cryptocurrencies, and therefore, the scope must be narrowed even further to include only Bitcoin synonyms, such as ``Bitcoin'' and ``BTC.''
Using this method, we gather 5,496,138 Bitcoin-related tweets generated within the real data period of the Bitcoin price. 
Table~\ref{dataStat} lists the statistical values for the dataset. 
\begin{table}[h!]
\centering
\caption{Statistical information of dataset}
%\vspace{0.2cm}
\begin{tabular}{c|c} 
\hline
\textbf{Definition} & \textbf{Value}   \\ 
\hline
%\hline
Time frame for prediction (start) & 01.04.2014 \\ 
%\hline
Time frame for prediction (end) & 14.11.2018\\ 
%\hline
Number of total tweets & 5,496,138 \\ 
%\hline
Tweet with keywords used once & 3,462,567\\ 
%\hline
Tweet with keywords used twice & 1,154,192\\
%\hline
Tweet with keywords used more than three time & 879,379\\
\hline
\end{tabular}\label{dataStat}
\end{table}

Tweets obtained directly from Twitter typically create noisy datasets. This is due to the social nature of social media use. Certain noises in tweets, such as URLs, emoticons, and user references, must be eliminated appropriately. For this purpose raw Twitter data must be formatted to build a dataset that can be easily processed by multiple classifiers. 
\begin{table}[h!]
%\vspace{-0.3cm}
\centering
\caption{Pre-processing tasks}
\begin{tabular}{c|c} 
\hline
\textbf{Number} & \textbf{Tasks} \\ 
\hline
%%\hline
1 & Change all letters in tweet to lower case \\ 
\hline
2 & Check and switch 2 or more dots (.) with space\\ 
\hline
3 & Switch 2 or more spaces with one single space \\ 
\hline
4 & Remove user-mentioning symbol (@)\\ 
\hline
5 & Change hashtags into typical words\\
\hline
6 & Remove retweet symbol (RT) and URLs\\
\hline
7 & Reduce characters repeated more than 3 times\\
\hline
\end{tabular}\label{preprocessing}
\end{table}
To this end, we consider several preprocessing steps to normalize the dataset, minimize its size, etc.
Table~\ref{preprocessing} presents an example of our preprocessing tasks, in which the above order is not important. We use the data refined according to the corresponding processing.

\subsection{Attribute Division}\label{formats}
To determine the effects of tweet attributes, we divide the preprocessed data into the following four types: (1) number of followers of the poster, (2) number of comments on the tweet, (3) number of likes, and (4) number of retweets. 

\smallskip
\noindent{\bf \em Sorting According to Attributes.} We consider that the tweet data covered tweets posted within 1,688 days, and we already obtain a single dataset with over 5 million tweets during this period.%, as indicated in Table~\ref{statistics}. 

%\begin{table}[ht]
%\vspace{-0.3cm}
%\centering
%\caption{Statistics of tweets sorted by posted date}
%\begin{tabular}{|p{150pt}|p{60pt}|} 
%\hline
%Total number of days & 1,688\bf{}  \\ 
%\hline
%Total number of tweets & 5,496,138\\ 
%\hline
%Maximum number of tweets posted in one day & 5,246 \\ 
%\hline
%Minimum number of tweets posted in one day & 1,742\\ 
%\hline
%Average number of tweets posted in one day & 3,256\\
%\hline
%\end{tabular}\label{statistics}
%\vspace{-0.1cm}
%\end{table}

To create datasets of interest, tweets posted on a particular day were separately sorted into four datasets according to the above attributes in decreasing order. That is, we sort the dataset by attribute (1), save it separately and sort it again by attribute (2), save it separately, and repeat this process with attributes (3) and (4). However, this is the same dataset.

\smallskip
\noindent{\bf \em Avoiding Similar Data.} To prevent similar data from appearing in each dataset, only the first half of each dataset is used in the experiment. In simple terms, all tweets posted in one day were sorted in decreasing order of their number of comments, and only the first half of the tweets were used as the first dataset. Subsequently, the tweets are disordered by the number of followers (1) and only the first half is used as the second dataset. Similarly, they are sorted according to the number of comments (2) and the number of retweets to create attributes (3) and (4).

\subsection{Sentiment Analysis}
As a final step, we apply sentiment analysis to determine the subjective emotions or views expressed in the tweets on Bitcoin.
We perform sentiment analysis by categorizing textual views into categories such as ``positive,'' ``negative,'' or ``neutral.'' We use the Valence Aware Dictionary and Sentiment Reasoner (VADER) \cite{Hutto2014} to classify the content of each tweet. VADER is a sentiment analysis Python library that uses lexicons and rules to analyze sentiments posted on social media. VADER includes three valence scores for each sentiment, given text content: positive, negative, and neutral. The valence ratings of each word in the lexicon are added together, modified according to the rules, and then normalized into $[-1,1]$, where $-1$ is extremely negative, $+1$ is extremely positive, and $0$ is neutral. These statistics are good because they provide a single unidimensional estimate of the emotion for each tweet. Based on this, we use the compound score to describe the sentiment of each tweet.
Subsequently, we perform proper Q-learning for the price prediction with sentimentally analyzed tweet data, as described in the following section. 

\section{Learning Algorithm}

In this section, we introduce our approach to predicting Bitcoin prices based on Twitter data. For this, we adopt simple reinforcement learning, in which the environment was the Bitcoin market. First, we briefly explain RL and the proposed approach with RL in the following subsection.  

\subsection{RL and Q-Learning}

Standard RL is formulated based on a Markov decision
Process (MDP). An MDP is a tuple $<\cS, \cA, r, P, \gamma>$, where
$\cS$ and $\cA$ are sets of states and actions, respectively, and
$\gamma \in [0, 1]$ denotes the discount factor. A transition probability
function $P : \cS \times \cA \rightarrow \cS$ maps the states and actions
to a probability distribution over the next states, and
$r : \cS\times\cA\rightarrow \real$ denotes the reward. The goal of RL
is to learn a policy $\pi: \cS \rightarrow \cA$ that solves the MDP by
maximizing expected discounted returns
$R_t =\expect{\sum_{k=0}^{\infty} \gamma^{k}r_{t+k}|\pi}$.  The policy
induces a value function $V^\pi(s)=\mathbb{E}_\pi[R_t|s_t=s]$ and an
action value function $Q^\pi(s,a)=\mathbb{E}_\pi[R_t|s_t=s,a_t=a]$.

In general, the state transition probability is often not provided in the RL. In this case, the agent must learn the optimal policy using trial and error through exploration. In RL, determining a policy that maximizes the expected reward through this process is known as model-free learning. Q-learning is one of the most famous model-free algorithms. RL strategies (such as Q-learning) have recently been used in various sectors to improve prediction models in various areas of social network research \cite{Yang2012}.
Q-learning \cite{watkins92} is a simple RL algorithm that provides the current state and finds the best action to be taken in that state. This is an off-policy algorithm because it learns from random actions. It constructs a Q-table $Q(s,a)$, where the value of the table is the reward when the agent selects action $a \in \cA$ at state $s \in \cS$. The algorithm operates in three basic steps: (1) the agent starts in a state, takes an action, and receives a reward; (2) for the next action, the agent has two choices: either reference the Q-table and select an action with the highest value, or take a random action; and (3) the agent updates the Q-values (i.e., $Q(s,a)$) in the table. The main objective is to learn the Q-function. To describe this precisely, let $s_t$ and $a_t$ be the state and action at current time $t.$ Before the iteration, Q is initialized to an arbitrary value. Subsequently, at each time $t,$ the agent selects an action $a_t$ at $s_t$ and observes a reward $r_t$, following which it enters a new state, $s_{t+1}$. Subsequently, the values of Q are updated. At the core of the algorithm is the Bellman equation as a simple value iteration update using the weighted average of the old value and new information: 
\begin{align}
  \label{eq:qlearning}
Q&^{new}(s_{t},a_{t})\cr
&=Q(s_{t},a_{t})+\theta * [r_{t}+\gamma Q^*(s_{t+1},a')-Q(s_{t},a_{t})],
\end{align}
where $\theta$ ($0 < \theta \leq 1$) is the learning rate and $\gamma$ is a discount factor with $0 \leq \gamma \leq 1$. The value of $Q^*$ is the estimate of the optimal future value, which is expressed by 
\begin{align}
  \label{eq:Q}
Q^*=\max_{a'}Q(s_{t+1},a').
\end{align}
This process continues until $s_{t+1}$ reaches its final or terminal state. Due to the lack of model information (the transition probability of the Bitcoin price), we adopt Q-learning as an RL approach for our Bitcoin price prediction problem.

\begin{figure}[t!]
\begin{center} \centering
\includegraphics[width=1\linewidth]{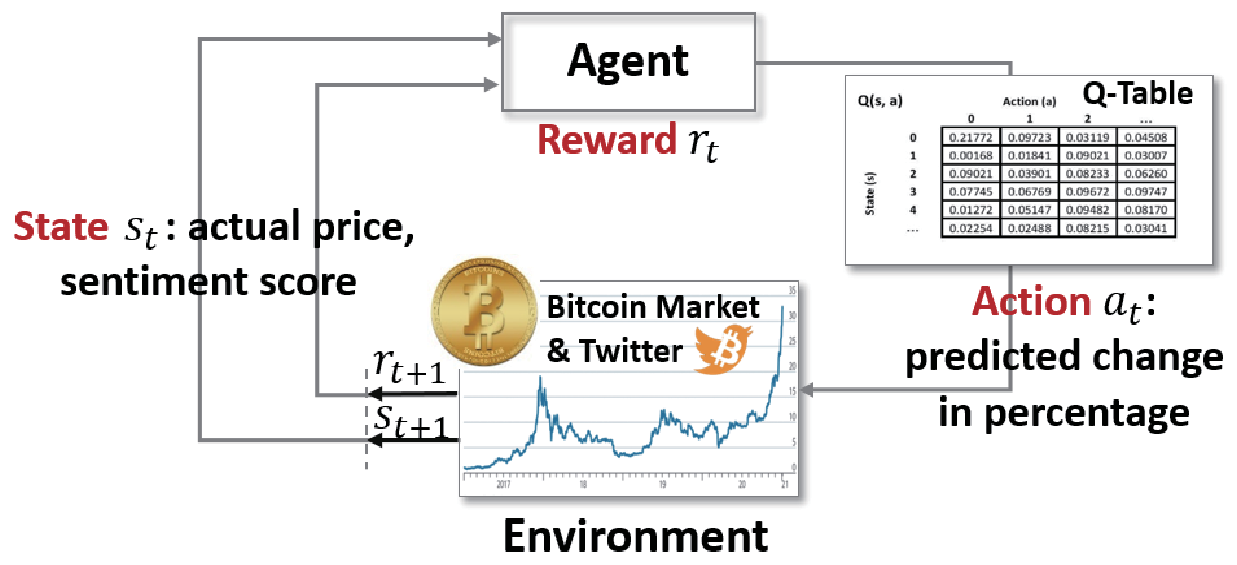}
\caption{Q-learning of our Bitcoin price prediction. }
\label{fig:Highway}
\vspace{-0.1cm}
\end{center}
\end{figure}

\subsection{Bitcoin Price Learning}
In this prediction problem, an agent interacts with the environment, which is the Bitcoin market, and learns how to predict future prices based on Q-learning. For this purpose, we define a tuple $<\cS, \cA, r>$, as follows:

\begin{itemize}
\item \noindent{\bf \em State Space} $\cS$. As a state $s_t:=(AP_t, TS_t) \in \cS$ of the agent at time $t,$ where $AP_t$ is the actual Bitcoin price and $TS_t$ is the tweet sentiment score at time $t$, respectively. The Bitcoin price is usually expressed with two decimal places (e.g., 21,254.50 USD.) and we consider the tweet sentiment score as a discrete value after applying round to two decimal places. Hence, we note that the considered state space is also discrete. 
\smallskip
\item \noindent{\bf \em Action Space} $\cA$. The action $a_t \in \cA$ of the agent at time $t$ is defined as a prediction of the current Bitcoin price. However, to reduce the number of action states, the percentage of the current price increasing, decreasing, or not changing is selected. That is, the action space is the rate of the price change as a percentage, which is expressed by $\cA:=\{-1000,-999,...,0,...,999,1000\}$\footnote{This is because it rarely increases or decreases by more than 1000\% compared with the previous day in Bitcoin price prediction.}. For example, if the agent selects 50, it means that the agent predicts the next price by increasing the current price by 50\% of the current price; that is, $a_t = 1.5 \times AP_{t-1}$. We also denote this action by the predicted price $PP_t$ at time $t.$

\smallskip
\item \noindent{\bf \em Reward Function} $r$. For the prediction of the actual Bitcoin price, we consider the following three reward functions: (1) simple difference reward (SDR), (2) relative difference reward (RDR), and (3) comparative difference reward (CDR). Detailed description for each function is as follows:

\begin{itemize}
    \item[$(i)$ \noindent{\bf \em SDR.}] This reward function is simply based on the difference between the actual price ($AP_{t}$) and the predicted price ($PP_{t}$). Considering that the model needs to receive a higher reward for a smaller difference, it receives only negative rewards with the highest possible reward $r_t =0$ in case that $AP_t$ and $PP_t$ are the same. Formally, the SDR is defined by: 
\begin{align}
  \text{SDR:}~\label{eq:r1}
r_t= -|AP_t-PP_t|.
\end{align}

\smallskip
    \item[$(ii)$ \noindent{\bf \em RDR.}] It is based on the relative difference between $AP_t$ and $PP_t$, which is formally defined by:
\begin{align}
  \text{RDR:}~\label{eq:r1}
r_t= \frac{-|AP_t-PP_t|}{AP_t}*100\%,
\end{align}
where $AP_t >0.$ Therefore, $r_t \in [-\infty,0]$ where $r_t =0$ means perfect fit of $PP_t$ to the $AP_t$.

\smallskip
    \item[$(iii)$ \noindent{\bf \em CDR.}] In the prediction of actual price of Bitcoin, it will be an important information on how much has increased or decreased compared to the previous step. In the third reward function, we consider the additional information about this rate of change. To formally describe this, we first introduce a concept of zero-reward value as follow. 
  \par\noindent\hrulefill
    \begin{definition}
\label{def:zero reward}
Let $\alpha = (AP_{t} - AP_{t-1})/AP_{t-1}$ where $AP_{t-1}>0$ \ie the rate of change of actual price. Let $l=AP_t - PP_{t-1}(1+\alpha)>0$. We call a point by  \emph{Zero-value reward (ZR)} where the difference from $AP_t$ is $l.$
\end{definition}
\vspace{-0.2cm}
  \par\noindent\hrulefill

\smallskip
    Actually, we have two such zero-reward values as shown in Figure~\ref{reward} since one point is less than $l$ from $AP_t$ and the other is larger than $l$ from it. We denote the former by $ZR_{t}^{1}$ and the latter by $ZR_{t}^{2}$, respectively. Then, $ZR_{t}^{1}$ is computed by (See Figure~\ref{reward}): 
     \begin{align}
        \text{$ZR_{t}^{1}$}~\label{zr1}
         = PP_{t-1} + ({PP_{t-1}*\alpha}),
        \end{align}
and the $ZR_{t}^{2}$ is computed by 
    \begin{align}
        \text{$ZR_{t}^{2}$}~\label{zr1}
         = PP_{t-1} + ({PP_{t-1}*\alpha}+2l).
        \end{align}

\begin{figure}[t!]
\begin{center} \centering
\includegraphics[width=1\linewidth]{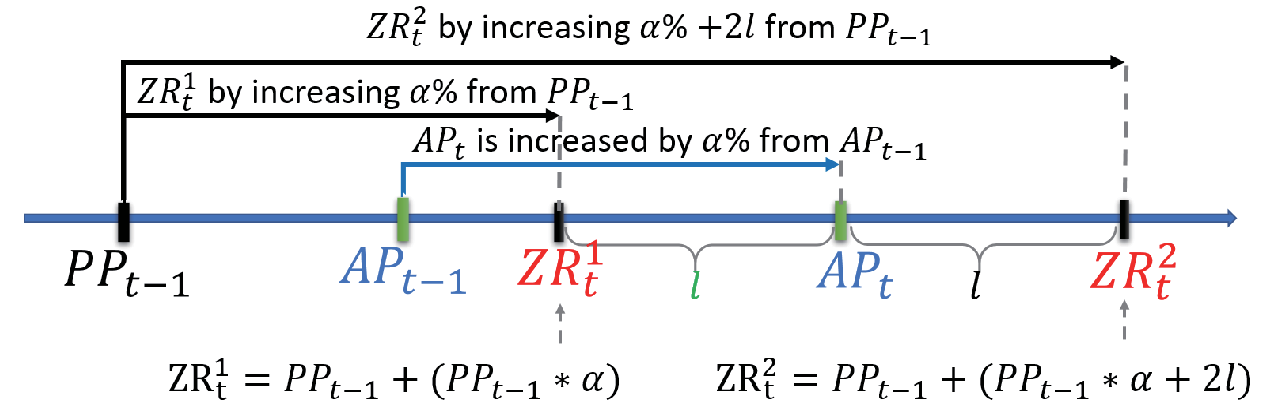}
\caption{Computation of two zero-valued rewards $ZR^1_{t}$ and $ZR^2_{t}$ used in the paper. ($AP_t$- actual price at $t$ time, $PP_t$- predicted price at time $t$.) }
\label{reward}
\vspace{-0.1cm}
\end{center}
\end{figure}
From two zero-reward points, we compute the reward value based on whether $PP_t$ is higher or lower than the $AP_{t}$. The formula of computing the reward value is different according to the value of $PP_t$. The explanation of the possible $PP_t$ cases and computing formulas is as follows.
    \begin{itemize}
        \item[$(a)$] The case where the $PP_{t}$ is smaller than the actual price ($PP_t<AP_t$). As $AP_{t}$ value stands in the middle of positive rewards interval; the agent receives a negative reward if $PP_t<ZR_{t}^{1}$ or a positive reward if the $PP_t$ is between $ZR_{t}^{1}$ and $AP_t$ ($ZR_{t}^{1} < PP_t\leqslant AP_t$). The value of the reward is calculated as follows:
        \begin{align}
        \text{$r_t$}~\label{case1}
         =\frac{PP_t-ZR_{t}^{1}}{AP_t-ZR_{t}^{1}} *100\%
        \end{align}
    
    \item[$(b)$] The case where the prediction price is higher than the $AP_{t}$ ($PP_t>AP_t$). As given in interval determination paragraph, when the $PP_{t}$ is higher than the $AP_{t}$, the model computes $ZR_{t}^{2}$ value to decide whether the reward is positive or negative. If the $PP_{t}$ is in between $AP_{t}$ and $ZR_{t}^{2}$ ($AP_{t}\leqslant PP_t < ZR_{t}^{2}$) value range, then the reward will be positive. If the $PP_{t}$ is higher than $ZR_{t}^{2}$, the reward will be negative. This is computed by: 
    
        \begin{align}
        \text{$r_t$}~\label{case2}
         =\frac{PP_t-ZR_{t}^{2}}{AP_t-ZR_{t}^{2}}*100\%
         \end{align}
    
    \end{itemize}

%The model calculates the difference between $ZR_1$ and $AP_t$, and defines the second zero-reward value ($ZR_2$) by adding the same difference to the $AP_t$. 
%If $PP_t$ is between the $ZR_1$ and $ZR_2$, then model receives the positive value reward up to $r_t =100$ (100 reward when $PP_t$=$AP_t$) based on how close the $PP_t$ to the $AP_t$.
%        \begin{align}
%        \label{eq:r1}
%        r_{t}=\left | \tfrac{PP_t-ZR_1}{AP_t-ZR_1} \right |*100\%
%        \end{align}
%If $PP_t$ is out of $ZR_1$ and $ZR_2$ range, then the model receives a negative reward based on how far the $PP_t$ from the $ZR_1$ (if $PP_t$ is greater than $ZR_2$ then it is based on how far from the $ZR_2$).
%    \begin{align}
%    \label{eq:r1}
%    r_{t}=-\left | \tfrac{PP_t-ZR_1}{AP_t-ZR_1} \right |*100\%
%    \end{align}

\end{itemize}
%In all cases, when the predicted price is closer to the actual price, the reward is higher (the highest reward is $r_t =0$ if they are the same). Unlike the first reward, the second reward, namely the RDR, also varies according to the current Bitcoin price. That is, even if the difference between the predicted and actual prices is the same, the reward value is higher if the current price is large.
In all the cases, when the predicted price comes closer to the actual price, the reward value becomes higher. The second reward function, namely the RDR, also varies according to the current Bitcoin price. That is, even if the difference between the predicted and actual prices is the same, the reward value is higher if the current price is large.  %All three proposed reward functions are shown visually in Figure \ref{reward}.

\end{itemize}

\smallskip
Based on the defined reward functions and preprocessed tweet data, the agent learns the actual Bitcoin price and attempts to make a prediction by repeating the following working steps:
\begin{itemize}
    \item Agent starts in a state ($s_1=(AP_1, TS_1)$ - Actual price of Bitcoin, sentiment score), takes an action ($a_1$ - a number between -1000 to 1000, as it will be applied as a percentage of change to actual price), and receives a reward ($r$ - computed based on one of SDR, RDR, and CDR reward functions).
    \item The agent chooses an action by referring to the highest value in Q-table.
    \item Update Q-values.
\end{itemize}
As the Q-values are updated and the agent chooses the maximum value in the table to take the action, the agent performance also starts to improve. The model with the above parameters is tested using four different datasets to experimentally verify the predictability range of tweet attributes. A brief explanation of the experiment is presented in section V. 

\vspace{-0.2cm}
\section{EXPERIMENT AND RESULTS}
%In this section, we present our experimental results for the prediction performance.
In this section, we present three different experimental results in order to determine the best reward function, tweet attribute that has the most influence on price, and computer resource working overloads during the performance of both classic and proposed approaches.
For this, we use Python to create the experimental environment and the Pandas library for data preprocessing. Sentiment analysis is performed using the VADER analyzer tool, and TensorFlow and Keras are used for training and testing, respectively. 
For monitoring and analyzing of computer resources (CPU, RAM, and memory) usage we use one of Windows 10 standard tools called Performance Monitor \cite{PerMon}. It is useful with its options where anyone can customize what data to collect, when the collection begins, how long the analysis process needs to run, etc. 

\smallskip
\noindent{\bf \em Training with Q-Learning.}
In the model training, we use a dataset of tweets posted between April 1, 2014, and June 30, 2017. 
The training process yielded promising results when the first part of the divided dataset was used to feed the model. We use $\gamma=0.95$ as the discount factor because this value provided the best performance during the experiment. 

\smallskip

\subsection{Performance Measures}
%To evaluate the performance of our model with reward functions, we use a variety of evaluation metrics as in Table~\ref{metrics}. 
To evaluate the performance of our model with reward functions, we use six metrics among a wide range of evaluation metrics, as they are the most suitable for the prediction task and provide a valuable evaluation. We briefly describe them as follow. 
%All mathematical formulas of the employed evaluation metrics are given in Table \ref{metrics} and a brief explanation of these metrics is as follows.  

\begin{itemize}
\item[$(i)$] \noindent{\bf \em Variance Accounted For (VAF).}
 VAD \cite{vaf} is used to verify the correctness of a model by comparing the real output with the predicted output. The values of VAF which is closed to 100\% indicate highly accurate prediction. With the definition of the actual price  - $AP$ and the predicted price as - $PP$, the formula of VAF is given by:
\begin{align}
    \text{$VAF$}~\label{vaf}
     =\left(1-\frac{var(AP-PP)}{var(AP)}\right)*100\%,
\end{align}
where $var(x)$ is the variance of $x$, which is computed by $var(x)=(\sum_{t=1}^{n}(x_t-\bar{x})^2) /n-1.$
% \begin{align}
%     \text{$var(x)$}~\label{var}
%      =\frac{\sum_{t=1}^{n}(x_t-\bar{x})^2}{n-1}.
% \end{align}
Here, $x_t$ is a value of $x$ at time $t$ and $\bar{x}$ is the average value of $x_t$ from $1 \leq t \leq  n$. In our experiment, we set $n$ = 1690 for all performance metrics.

\smallskip
\item[$(ii)$] \noindent{\bf \em Coefficient of Determination ($R^2$).}
$R^2$ is used to evaluate the forecast outputs and provides a measure of how well-observed outcomes are replicated by the model \cite{r2}. Formally, it is computed by:
\begin{align}
    \text{$R^2$}~\label{r2}
     =1-\frac{RSS}{TSS}.
\end{align}
where RSS is the residual sum of squares which is given by $RSS=\sum_{t=1}^{n}(AP_{t}-PP_{t})^2$ and TSS is the total sum of squares that is $TSS=\sum_{t=1}^{n}(AP_{t}-\overline{AP})^2.$ Here, $AP_t$ and $PP_t$ denote the actual price and predicted price of Bitcoin at time $t,$ and $\overline{AP}$ is the average value of $AP_t$ for time $1 \leq t \leq n$, respectively. 
Hence, the range of $R^2$ is $[0,1]$, where 1 indicates a perfect match of the prediction data with actual data. 
%In the aforementioned formulas, the $t$ value indicates the specific time slot, the $n$ value denotes the total number of experimental time slots, where in this study its value equals $n$ = 1690, and $\bar{AP}$ describes the average actual price.  
%\begin{table}[]
%\centering
%\caption{Summary of evaluation metrics in the paper. ($AP_t$ - actual price at time $t$, $PP_t$ - predicted price  at time $t$, $\overline{AP}$ - average of actual price.)}
%\vspace{0.2cm}
%\begin{tabular}{c|c}
% \begin{tabular}{M{0.3\linewidth}|M{0.6\linewidth}}
%\hline
%\textbf{Evaluation \\metrics} & \textbf{Equation} \\ 
%\hline
%VAF & \thead{$\left(1-\frac{var(AP-PP)}{var(AP)}\right)*100\%$} \\ 
%\hline
%$R^2$ & \thead{$R^2=1-(RSS/TSS)$ \\$RSS=\sum_{t=1}^{n}(AP_{t}-PP_{t})^2,$\\ $TSS=\sum_{t=1}^{n}(AP_{t}-\overline{AP})^2$} \\ \hline
%MAPE & \thead{$\frac{1}{n}\sum_{t=1}^{n}\left | \frac{AP_{t}-PP_{t}}{AP_{t}} \right |*100\%$} \\ 
%\hline
%NSE & \thead{$1-\frac{\sum_{t=1}^{n}(AP_{t}-PP_{t})^2}{\sum_{t=1}^{n}(AP_{t}-\overline{AP})^2}$}\\ \hline
%RMSE & \thead{$\sqrt{\frac{\sum_{t=1}^{n}(AP_{t}-PP_{t})^2}{n}}$}\\
%\hline
%WMAPE & \thead{$\frac{\sum_{t=1}^{n}\left | AP_{t}-PP_{t} \right |}{\sum_{t=1}^{n}AP_{t}}*100\%$}\\
%\hline
%\end{tabular}\label{metrics}
%\end{table}

\begin{table}
\centering
\refstepcounter{table}
\caption{Summary of evaluation metrics in the paper. ($AP_t$ - actual price at time $t$, $PP_t$ - predicted price  at time $t$, $\overline{AP}$ - average of actual price.)}
\begin{tabular}{c|l} 
\hline
\begin{tabular}[c]{@{}c@{}}\textbf{Evaluation~}\\\textbf{metrics}\end{tabular} & \multicolumn{1}{c}{\textbf{Equation}}\\ \hline
VAF & \begin{tabular}[c]{@{}l@{}} \\$\left(1-\frac{var(AP-PP)}{var(AP)}\right)*100\%$\\[10px]\end{tabular}  \\ \hline
$R^2$ & \begin{tabular}[c]{@{}l@{}} \\$R^2=1-(RSS/TSS)$\\$RSS=\sum_{t=1}^{n}(AP_{t}-PP_{t})^2$\\$TSS=\sum_{t=1}^{n}(AP_{t}-\overline{AP})^2$\\[10px]\end{tabular}  \\ \hline
MAPE  & \begin{tabular}[c]{@{}l@{}} \\$\frac{1}{n}\sum_{t=1}^{n}\left | \frac{AP_{t}-PP_{t}}{AP_{t}} \right |*100\%$\\[10px]\end{tabular} \\ \hline
NSE & \begin{tabular}[c]{@{}l@{}} \\$1-\frac{\sum_{t=1}^{n}(AP_{t}-PP_{t})^2}{\sum_{t=1}^{n}(AP_{t}-\overline{AP})^2}$\\[10px]\end{tabular}  \\ \hline
RMSE  & \begin{tabular}[c]{@{}l@{}} \\$\sqrt{\frac{\sum_{t=1}^{n}(AP_{t}-PP_{t})^2}{n}}$\\[10px]\end{tabular}  \\ \hline
WMAPE  & \begin{tabular}[c]{@{}l@{}} \\$\frac{\sum_{t=1}^{n}\left | AP_{t}-PP_{t} \right |}{\sum_{t=1}^{n}AP_{t}}*100\%$\\[10px]\end{tabular} \\ \hline
\end{tabular}
\end{table}

\smallskip
\item[$(iii)$] \noindent{\bf \em Mean Absolute Percentage Error (MAPE).} 
Like the aforementioned metrics, MAPE is also used to measure the prediction accuracy but unlike them, it is commonly used as a loss function in model evaluation because of its highly intuitive interpretation in terms of relative error \cite{mape}. The formal computation is given by:
\begin{align}
    \text{$MAPE$}~\label{mape}
     =\frac{1}{n}\sum_{t=1}^{n}\left | \frac{AP_{t}-PP_{t}}{AP_{t}} \right | *100\%.
\end{align}

\smallskip
\item[$(iv)$] \noindent{\bf \em Nash–Sutcliffe model efficiency coefficient (NSE).}
The fourth evaluation metric that we consider to use is NSE, which is used to assess the predictive skill of models \cite{nse}. Following formula used to calculate the NSE value of the model prediction.
\begin{align}
    \text{$NSE$}~\label{nse}
     =1-\frac{\sum_{t=1}^{n}(AP_{t}-PP_{t})^2}{\sum_{t=1}^{n}(AP_{t}-\bar{AP})^2}.
\end{align}
Hence, the NSE becomes one in the case of a perfect prediction.

\smallskip
\item[$(v)$] \noindent{\bf \em Root-mean-square error (RMSE)}
This evaluation metric is frequently used to measure the difference between values predicted by the model and observed values \cite{rmse}. Formally, it is computed by: 
\begin{align}
    \text{$RMSE$}~\label{rmse}
     =\sqrt{\frac{\sum_{t=1}^{n}(AP_{t}-PP_{t})^2}{n}}.
\end{align}
Hence, we see that RMSE value is always non-negative, and a lower RMSE indicates a more accurate prediction than a higher RMSE. 

\smallskip

\item[$(vi)$] \noindent{\bf \em Weighted Mean Absolute Percentage Error (WMAPE)}
WMAPE is a variant of MAPE in which errors are weighted by values of actuals \cite{wmape}. The advantage of this metric over MAPE is that it overcomes the ``infinite error" issue \cite{wmape2}. The formal metric is defined by:
\begin{align}
    \text{$WMAPE$}~\label{wmape}
     =\frac{\sum_{t=1}^{n}\left | AP_{t}-PP_{t} \right |}{\sum_{t=1}^{n}AP_{t}}*100\%.
\end{align}
\end{itemize}

%\begin{figure*}[t!]
%\begin{center}
%\subfigure[SDR]{\includegraphics[width=0.672\columnwidth]{Figures/SDR.eps}\label{fig:result1}}
%\subfigure[RDR]{\includegraphics[width=0.672\columnwidth]{Figures/RDR.eps}\label{fig%:result2}}
%\subfigure[Posted Time]{\includegraphics[width=0.672\columnwidth]{Figures/posted.eps%}\label{fig:result3}}
%\vspace{-0.3cm}
%\end{center}
%\vspace{-0.2cm}
%\caption{Prediction performance of Bitcoin price for four attributes with two reward functions ((a) SDR and (b) RDR) for testing datasets. (AP: Actual Price, PP($x$): Predicted Price with attribute $x.$) (c). Prediction results of posted time attribute (daytime and nighttime).}
%\label{fig:result}
%\end{figure*}

\subsection{Results for each reward function with four attributes}

\begin{figure}[t!]
\begin{center}
\subfigure[SDR]{\includegraphics[width=0.9\columnwidth]{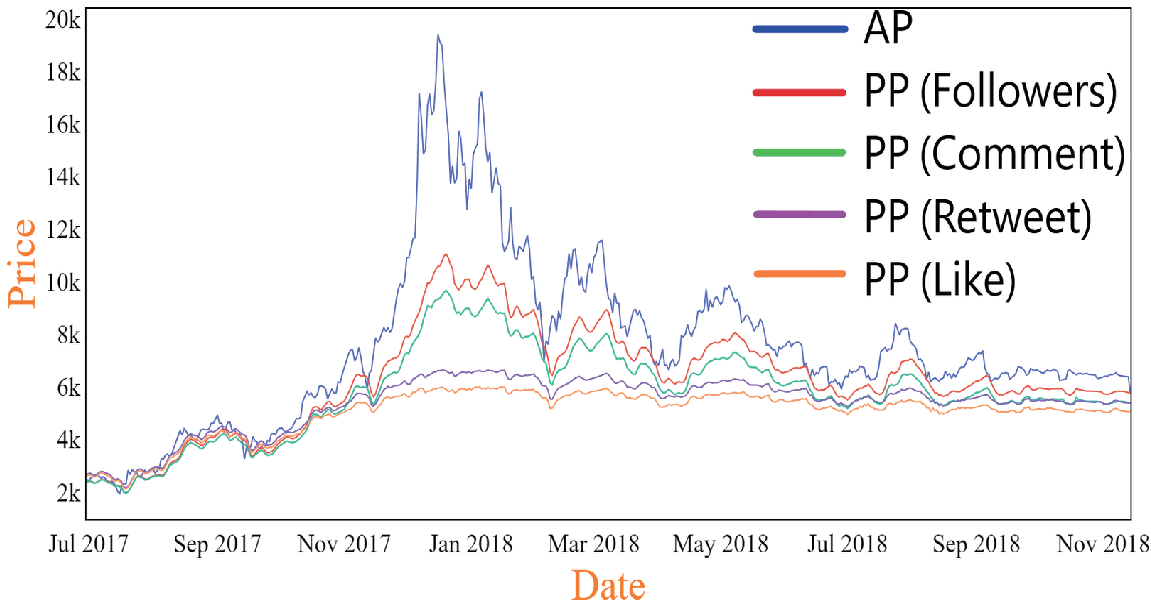}\label{fig:result1}}
\subfigure[RDR]{\includegraphics[width=0.9\columnwidth]{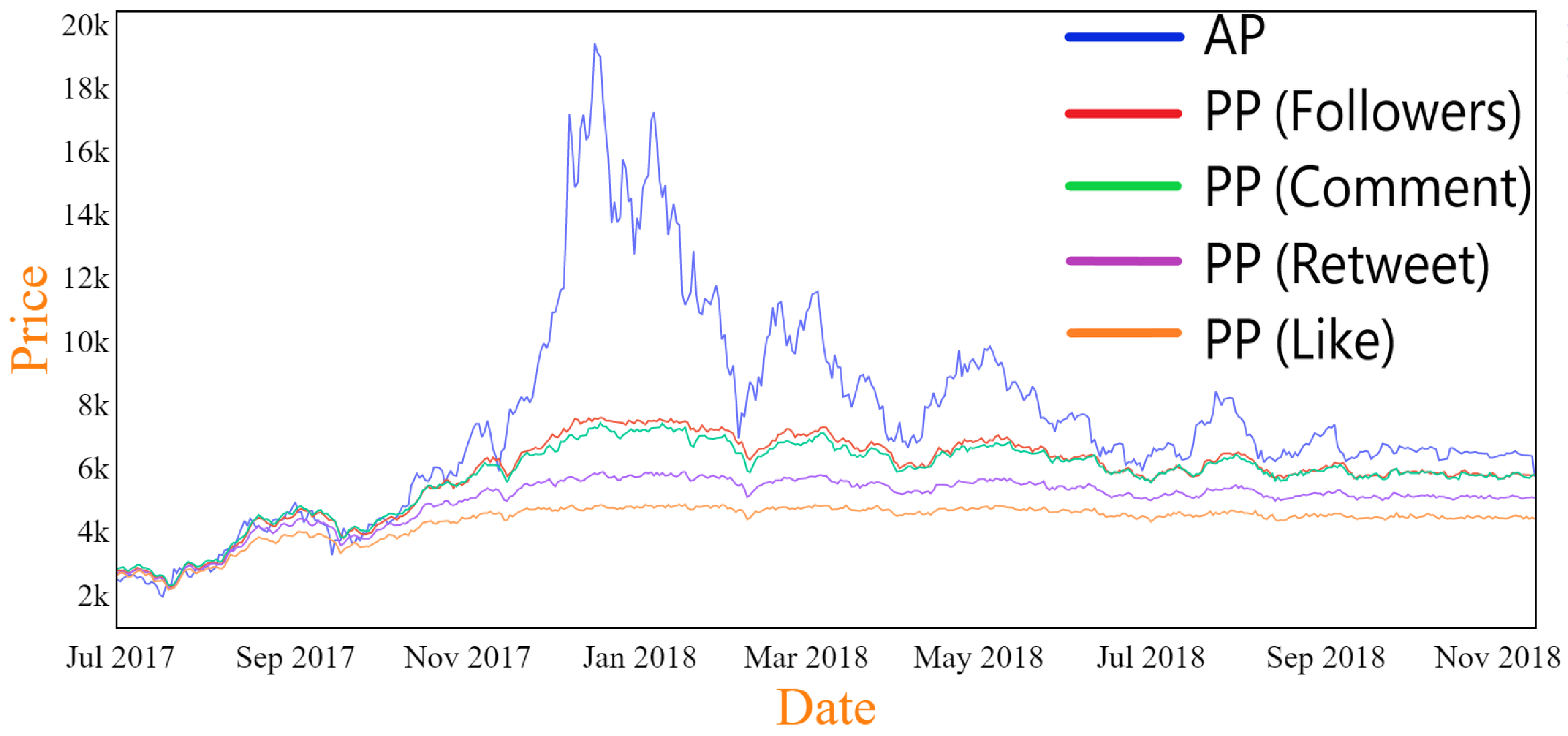}\label{fig:result2}}
\subfigure[CDR]{\includegraphics[width=0.9\columnwidth]{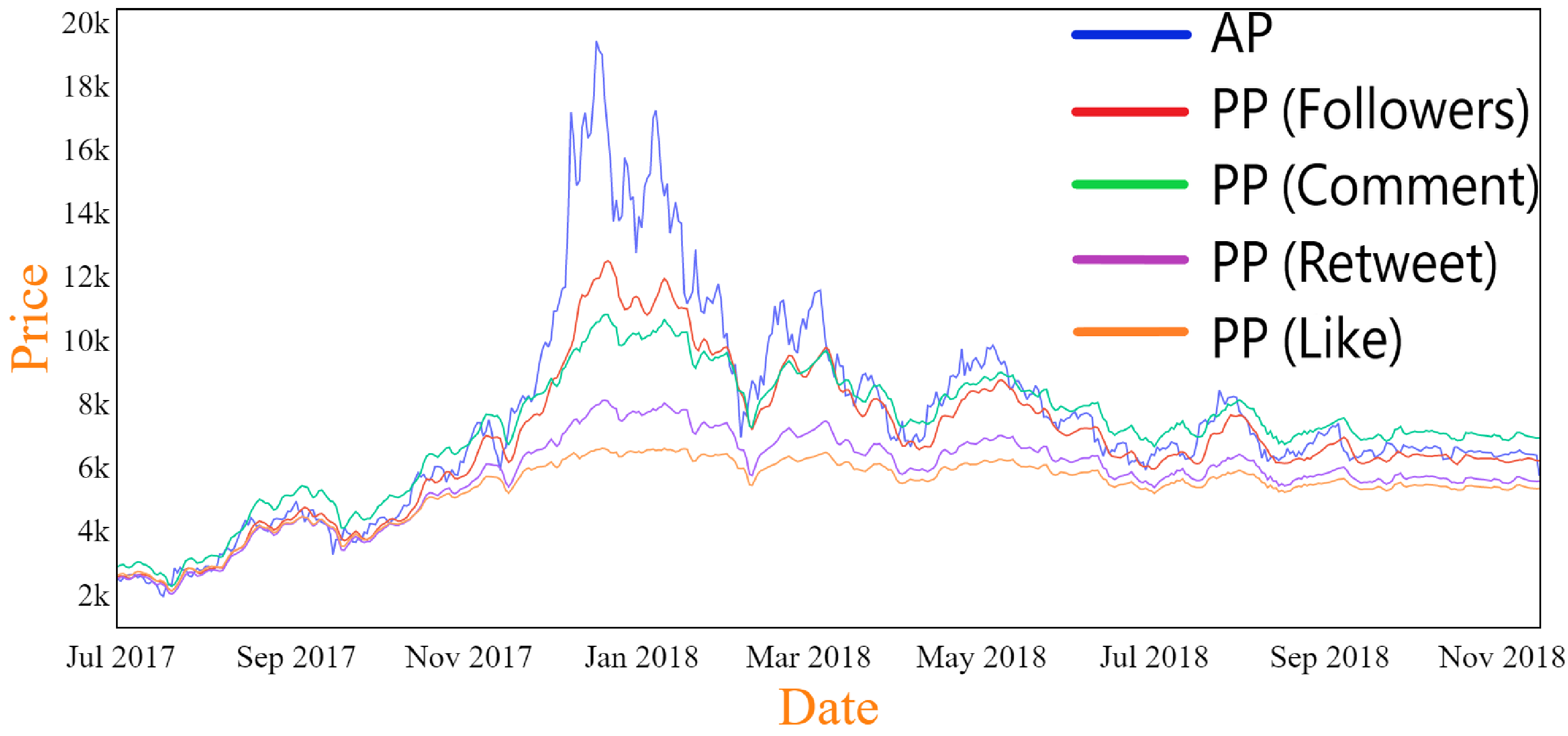}\label{fig:result3}}
\end{center}
\caption{Prediction performance of Bitcoin price for four attributes with three reward functions ((a) SDR, (b) RDR, and (c) CDR) for testing datasets. (AP: actual price, PP($x$): predicted price with attribute $x.$)}
\label{fig:result}
\end{figure}

As a first experiment result, we will show the prediction performance for the three reward functions to determine the most useful tweet attribute in predicting the price.
For this, we use a dataset of tweets posted between July 1, 2017, and November 14, 2018. We obtain the prediction results based on four attributes: most commented, most liked, most retweeted, and the number of poster followers. 

\smallskip
\bnote{\noindent{\bf \em Tweet Attribute Classification.} }
First, in Figure~\ref{fig:result}, we see that tweets posted by those with the most followers and tweets with the most comments exhibit the best prediction results for all the SDR, RDR, and CDR reward functions. However, the prediction with CDR is better than that with SDR and RDR because the CDR provides a reward by comparing the current action $a_t$ with the previous action $a_{t-1}$. Each action comparison with the previous action provides the opportunity to compare all actions relative to each other which boosts the learning process. The result shows that there is a high chance that people's tweets with the most followers catch the public's attention by being viral and have some influence on future events.
Moreover, it can be seen from the results of the experiment, that there is a ranking among the attributes based on their predictive powers. Among three prediction outputs with the three different reward functions, the dataset sorted by the number of user followers shows the most accurate prediction. Next, the dataset created from tweets with the most comments shows a more accurate forecast than the remaining two datasets. As the prediction results in Figure \ref{fig:result}, the most retweeted attribute comes in third place, whereas the most liked attribute is in the last place. 
\begin{table}[t!]
\label{tab1}\centering 
\caption{Performance metrics for prediction with SDR, RDR, and CDR.}
%\vspace{0.2cm}
\begin{tabular}{c}
\centering \textbf{With SDR}
\vspace{0.1cm}
\end{tabular}
\begin{tabular}{|p{35pt}|c|c|c|c|c|c|} 
\hline
\centering & \textbf{VAF} & \ $R^2$  & \textbf{MAPE}  & \textbf{NSE} & \textbf{RMSE}  & \textbf{WMAPE}  \\ \hline
\centering \textbf{Follower} & 78.85 & 0.63 & 13.919 & 0.62 & 1944.3 & 16.9\\ \hline
\centering \textbf{Comment} & 71.84 & 0.43 & 19.257 & 0.44 & 2421.3 & 22.9\\ \hline
\centering \textbf{Retweet} & 42.79 & 0.32 & 22.833 & 0.32 & 3243.7 & 29.1\\ \hline
\centering \textbf{Like} & 36.31 & 0.22 & 26.785 & 0.23 & 3574.2 & 33.3\\ \hline
%\hline
%\centering \textbf{Daytime posted} & 83.95 & 0.40 & 16.045 & 0.42 & 2495.3 &0.208\\ 
%\hline
%\centering \textbf{Nighttime posted} & 79.20 & 0.11 & 20.794 & 0.12 & 3034.7 &0.266\\ 
%\hline
\end{tabular}\label{SDRMetrics}
\vspace{0.2cm}

\begin{tabular}{c}
\centering \textbf{With RDR}
\vspace{0.1cm}
\end{tabular}
\begin{tabular}{|p{35pt}|c|c|c|c|c|c|} \hline
\centering & \textbf{VAF} & \ $R^2$  & \textbf{MAPE}  & \textbf{NSE} & \textbf{RMSE}  & \textbf{WMAPE}  \\ \hline
\centering \textbf{Follower} & 51.87 & 0.25 & 17.690 & 0.26 & 2789.5 & 23.1\\ \hline
\centering \textbf{Comment} & 47.41 & 0.18 & 18.948 & 0.18 & 2925.8 & 24.5\\ \hline
\centering \textbf{Retweet} & 33.13 & 0.10 & 27.408 & 0.11 & 3651.9 & 34.1\\ \hline
\centering \textbf{Like} & 22.96 & 0.06 & 35.439 & 0.06 & 4250.5 & 42.6\\ \hline
%\hline
%\centering \textbf{Daytime posted} & 76.03 & 0.41 & 23.969 & 0.41 & 3246.4 &0.299\\ 
%\hline
%\centering \textbf{Nighttime posted} & 65.248& 0.36 & 34.751 & 0.36 & 4104.6 &0.413\\ 
%\hline
\end{tabular}\label{RDRMetrics}
\vspace{0.2cm}

\begin{tabular}{c}
\centering \textbf{With CDR}
\vspace{0.1cm}
\end{tabular}
\begin{tabular}{|p{35pt}|c|c|c|c|c|c|} \hline
\centering & \textbf{VAF} & \ $R^2$  & \textbf{MAPE}  & \textbf{NSE} & \textbf{RMSE}  & \textbf{WMAPE}  \\ \hline
\centering \textbf{Follower} & 84.81 & 0.80 & 8.450 & 0.81 & 1441.2 & 10.7\\ \hline
\centering \textbf{Comment} & 71.87 & 0.70 & 11.510 & 0.69 & 1743.1 & 13.3\\ \hline
\centering \textbf{Retweet} & 57.68 & 0.34 & 19.078 & 0.33 & 2764.7 & 24.2\\ \hline
\centering \textbf{Like} & 42.64 & 0.26 & 23.378 & 0.27 & 3281.7 & 29.6\\ \hline
\end{tabular}\label{CDRMetrics}
\end{table}

\smallskip
\noindent{\bf \em Performance Evaluations.} 
To see the sufficient prediction performance, we obtain six different evaluation metrics during the assessment of performance for each reward function and each attribute are listed in Table \ref{SDRMetrics}.
First, we see that, in the case of CDR, the VAF values show the most accurate prediction compared with SDR and RDR. Further, the attribute of number of poster's followers has the highest prediction performance as we expected. 

% For example, the first attribute (number of poster's followers) has a 84.81 value with CDR, while this attribute has 78.85 and 51.87 values with SDR and RDR, respectively. The second attribute (number of comments) also has the highest value of 71.87, whereas in the other two functions, they are equal to 71.84 and 47.41. The same comparison priority is maintained in the third attribute (number of retweets) with values of 57.68, 42.79, and 33.13 values. The last attribute (number of likes), despite showing the lowest values of VAF, still shows 42.64 accuracy with CDR, which is best compared with 36.31 and 22.96 values when the model applied SDR and RDR functions, respectively. 
%Logically, daytime posted tweets show a better result with 83.95\% rather than nighttime posted tweets: 79.2\%.

In contrast to VAF, the $R^2$ takes values in the range $[0,1]$ where 1 indicates an ideal prediction. Keeping this definition in mind and by comparing the $R^2$ values of each reward function, we can determine that the model achieves a more precise prediction with CDR by having a maximum 0.8 value rather than SDR and RDR by having 0.63 and 0.25, respectively. The maximum $R^2$ values are achieved with the dataset that consists of posters' tweets with the highest number of followers.

By scoping the three prediction outputs with metric MAPE, we obtain a result that indicates the level of error in the predictions. Therefore, a lower MAPE value indicates higher accuracy. The MAPE value also shows no contradiction in the priority of the CDR over the SDR and RDR functions. For example, while SDR is being implemented by the model, the first attribute has a value of 13.919, which is the lowest among the second, third, and fourth attributes, with 19.257, 22.833, and 26.785 values, respectively. During the RDR implementation, the model has the lowest prediction quality. The MAPE value of the first attribute increased to 17.690 in this scenario, but still dominates the remaining attributes.

For the NSE metric, we observe similar results as the $R^2$ metric. Because the performance values are quite similar, we refrained from analyzing the reward functions' preferability and ranking of attributes.% based on predictive power.

In using the RMSE, taking into account the fact that RMSE measurement is based on errors, a low value of RMSE indicates a more accurate prediction than a high value RMSE. While SDR is implementing by the model, the follower attribute has the lowest value among all attributes. The model has the poorest prediction quality during RDR implementation. In this case, the RMSE value of the follower attribute increased to 2789.5, but it still dominates the remaining attributes. As we expected, the RMSE also shows the best prediction when the model used CDR as a reward function. 

The WMAPE is the last evaluation metric used in this study. Because it is a variant of MAPE, a smaller WMAPE value indicates an accurate prediction. With respect to CDR, WMAPE values indicate the most accurate forecast when compared to SDR and RDR. For example, the first attribute has a CDR value of 10.7, although this attribute has SDR and RDR values of 16.9 and 23.1, respectively.

\smallskip
\noindent{\bf \em Performance Comparisons.} In order to detect how good the model's performance is, we compare the accuracy of our prediction along with other similar studies that used different approaches to achieve an accurate prediction. However, there are several problems that resist making a fair comparison: types of data and its time period are different across studies; the model design and its implementation are not explained in detail in some studies; diversity of the metrics that are used to evaluate the model's performance; and difficulties on gathering all source codes and run in the same PC environment. Therefore, in Table \ref{comparative_table}, we briefly compare the results of previous relevant work with our proposed method. Most references are used Twitter as the main data source to obtain Bitcoin price predictions and yet only a few of them have considered analyzing the PC resource usage level. In the Table, we use the terms as follows: $(i)$ Non-filtered: BTC historical price data is used as the main dataset without being filtered by any conditions and is used entirely in its form. $(ii)$ Non-attribute filtered: Bitcoin-related tweets are used as the main dataset without being filtered by any Twitter attribute and are used entirely in its form. $(iii)$ Attribute-filtered: Bitcoin-related tweets are used as the main dataset and the dataset has been used after filtering by the "number of followers" attribute. We see that the result in Ye \etal\cite{Ye2022} shows the highest accuracy level with an 88.74\% value but the resource usage did not considered. We observe that our proposed Q-learning model that considers only Bitcoin-related tweets that are posted by posters who have the most number of followers, considers the PC resource usage level while obtaining 84.81\% accuracy which overcomes most of the previous studies results. 

\begingroup
\setlength{\tabcolsep}{1.65pt} % Default value: 6pt
\renewcommand{\arraystretch}{2.0} % Default value: 1
\begin{table}
\centering
\caption{Performance comparison with other results.}
\label{comparative_table}
\begin{tabular}{c||ccccc} 
\hline
\textbf{Studies}& \textbf{Datasets}& \textbf{Filter}& \textbf{Model}& \textbf{Accuracy} & 
\begin{tabular}[c]{p{0.06\textwidth}}\centering\textbf{PC resource analysis}\end{tabular}\\ 

\hline
\begin{tabular}[c]{p{0.09\textwidth}}Mittal \etal\cite{Mittal2019} - 2019\end{tabular}& 
\begin{tabular}[c]{p{0.07\textwidth}}\centering BTC price, Google trends\end{tabular} &
\begin{tabular}[c]{p{0.05\textwidth}}non-filtered\end{tabular} &
\begin{tabular}[c]{p{0.03\textwidth}}LSTM, RNN\end{tabular} & 
\begin{tabular}[c]{p{0.09\textwidth}}\centering 66.66\%\end{tabular}&
\begin{tabular}[c]{p{0.06\textwidth}}\centering No\end{tabular}\\ \hline

\begin{tabular}[c]{p{0.09\textwidth}}Pant \etal \cite{Pant2018} - 2018\end{tabular} & 
\begin{tabular}[c]{p{0.07\textwidth}}\centering BTC price, Tweets\end{tabular} & 
\begin{tabular}[c]{p{0.05\textwidth}}non-attribute filtered\end{tabular} &
\begin{tabular}[c]{p{0.03\textwidth}}RNN\end{tabular} & 
\begin{tabular}[c]{p{0.09\textwidth}}\centering 77.62\%\end{tabular}&
\begin{tabular}[c]{p{0.06\textwidth}}\centering No\end{tabular}\\ \hline

\begin{tabular}[c]{p{0.099\textwidth}}McNally \etal \cite{McNally2018} - 2018\end{tabular} & \begin{tabular}[c]{p{0.07\textwidth}}\centering BTC price\end{tabular} & 
\begin{tabular}[c]{p{0.05\textwidth}}non-filtered\end{tabular} &
\begin{tabular}[c]{p{0.03\textwidth}}RNN, LSTM\end{tabular} & 
\begin{tabular}[c]{p{0.09\textwidth}}\centering 52.78\%\end{tabular}&
\begin{tabular}[c]{p{0.06\textwidth}}\centering Yes\end{tabular}\\ \hline

\begin{tabular}[c]{p{0.099\textwidth}}Ye \etal  \cite{Ye2022} - 2022\end{tabular} & 
\begin{tabular}[c]{p{0.07\textwidth}}\centering BTC price, Tweets\end{tabular} & 
\begin{tabular}[c]{p{0.05\textwidth}}non-attribute filtered\end{tabular} &
\begin{tabular}[c]{p{0.03\textwidth}}LSTM, GRU\end{tabular} & 
\begin{tabular}[c]{p{0.09\textwidth}}\centering 88.74\%\end{tabular}&
\begin{tabular}[c]{p{0.06\textwidth}}\centering No\end{tabular}\\ \hline

\begin{tabular}[c]{p{0.09\textwidth}}Gurrib \etal \cite{Gurrib2021} - 2021\end{tabular} & 
\begin{tabular}[c]{p{0.07\textwidth}}\centering BTC price, Tweets\end{tabular} & 
\begin{tabular}[c]{p{0.05\textwidth}}non-attribute filtered\end{tabular} &
\begin{tabular}[c]{p{0.03\textwidth}}LDA\end{tabular} & 
\begin{tabular}[c]{p{0.09\textwidth}}\centering 82.8\%\end{tabular}&
\begin{tabular}[c]{p{0.06\textwidth}}\centering No\end{tabular}\\ \hline

\begin{tabular}[c]{p{0.099\textwidth}}Sumarsih \etal \cite{Sumarsih2018} - 2018\end{tabular} & \begin{tabular}[c]{p{0.07\textwidth}}\centering BTC price\end{tabular} & 
\begin{tabular}[c]{p{0.05\textwidth}}non-filtered\end{tabular} &
\begin{tabular}[c]{p{0.03\textwidth}}LR\end{tabular} & 
\begin{tabular}[c]{p{0.09\textwidth}}\centering 73.15\%\end{tabular}&
\begin{tabular}[c]{p{0.06\textwidth}}\centering Yes\end{tabular}\\ \hline

\begin{tabular}[c]{p{0.09\textwidth}}Jain \etal\cite{Jain2018} - 2018\end{tabular}&
\begin{tabular}[c]{p{0.07\textwidth}}\centering BTC price, Tweets\end{tabular} & 
\begin{tabular}[c]{p{0.05\textwidth}}non-attribute filtered\end{tabular} &
\begin{tabular}[c]{p{0.03\textwidth}}MLR\end{tabular} & 
\begin{tabular}[c]{p{0.09\textwidth}}\centering 44\%\end{tabular}&
\begin{tabular}[c]{p{0.06\textwidth}}\centering No\end{tabular}\\ \hline

\begin{tabular}[c]{p{0.075\textwidth}}\bf{This paper}\end{tabular} & 
\begin{tabular}[c]{p{0.073\textwidth}}\centering \bf{BTC price, Tweets}\end{tabular} &
\begin{tabular}[c]{p{0.05\textwidth}}\bf{attribute \quad filtered}\end{tabular} &
\begin{tabular}[c]{p{0.03\textwidth}}\centering \bf{Q-learning}\end{tabular} & 
\begin{tabular}[c]{p{0.09\textwidth}}\centering \bf{84.81\%}\end{tabular}&
\begin{tabular}[c]{p{0.06\textwidth}}\centering \bf{Yes}\end{tabular}\\ \hline

\end{tabular}
\end{table}

% After looking through each metric's information, for further experimentation, we choose to use the CDR as a reward function and a dataset created based on users' tweets with the highest number of followers for the predictive model.

\subsection{Results with computer resource usage}
\bnote{
In this subsection, we will describe the comparison results between our proposed approach and the classic approach. These two are explained as follows.
\begin{itemize}
\item[$(a)$] \noindent{\bf \em Proposed approach:} In the proposed approach, we obtain the Bitcoin-related tweets only from those who have the most followers, \ie we use the data with attribute-filtering.
\item[$(b)$] \noindent{\bf \em Classic approach:} In this approach, we obtain all Bitcoin-related tweets, \ie we use all of the data without attribute-filtering.
\end{itemize}
For this, we perform two different experiments as follows. 
\begin{itemize}
\item[$(i)$] \noindent{\bf \em Fixed running time:} In the first experiment, we see how the resource usage and accuracy for each approach are different when the running time of same PC is equal to 1 hour.
\item[$(ii)$] \noindent{\bf \em Fixed target accuracy:} In the second experiment, we check how much the performance difference are when the target accuracy of prediction is fixed for both approaches. During the experiment, we observe the status of the CPU workloads, RAM, and memory usage.
\end{itemize}
The resource usage information is given using three types of metrics: minimum, average, and maximum values during the experiment. At the end of the experiment, we calculate the accuracy of both approaches and a comparison of the observed results is presented in Table \ref{Exp_2}.}

\begin{figure*}[t!]
\begin{center}
\subfigure[Resources usage graph with classic approach]{\includegraphics[width=1.0\columnwidth]{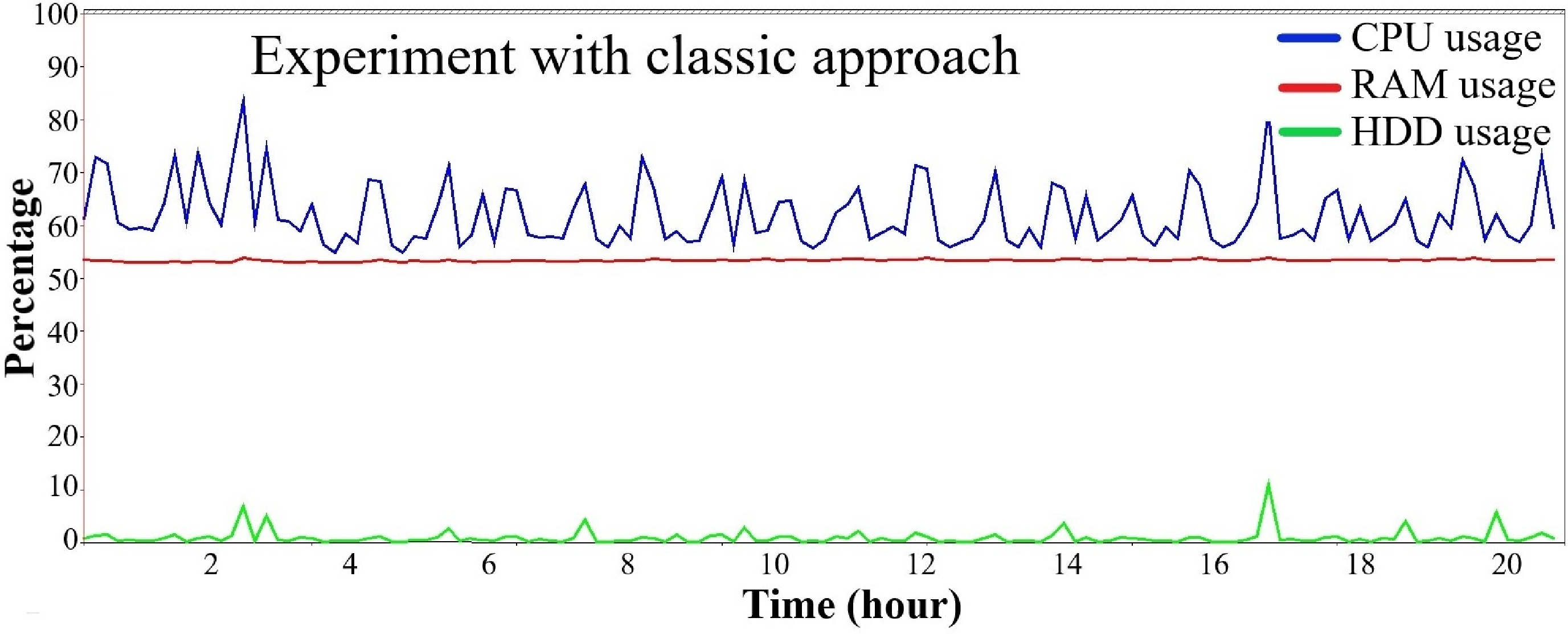}\label{21hour}}
\subfigure[Resources usage graph with proposed approach]{\includegraphics[width=1.0\columnwidth]{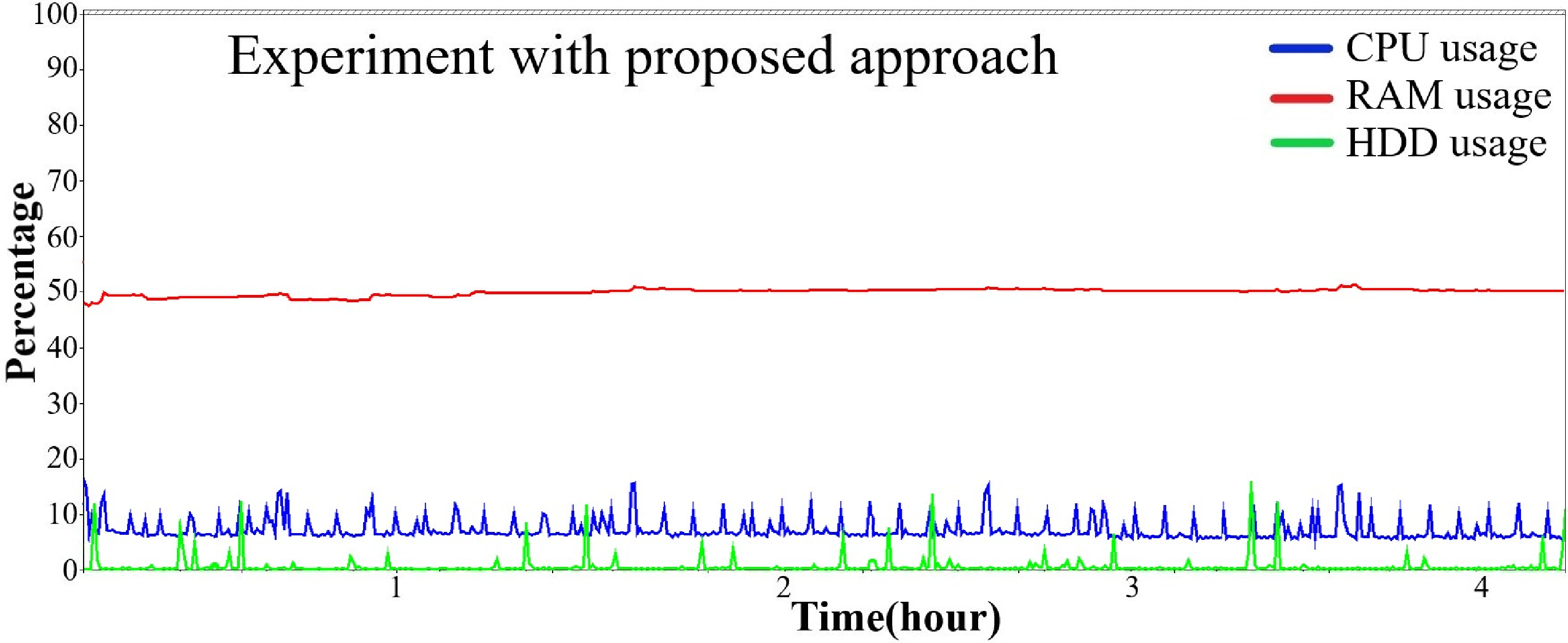}\label{4hour}}

\end{center}
\caption{Computer resources usage results for the two approaches (a) Classic approach and (b) Proposed approach. In the experiment, we set the target accuracy by VAF = 84.81\%. The blue line graph represents CPU workloads, red line indicates RAM usage, and green line describes memory usage.}
\label{Exp_3_Fig}
\end{figure*}

\begingroup
\setlength{\tabcolsep}{4pt} % Default value: 6pt
\renewcommand{\arraystretch}{1.5} % Default value: 1
\begin{table}[t!]
\label{tab1}\centering 
\caption{Experiment results of classic and proposed approaches}
%\vspace{0.2cm}
\begin{tabular}{c}
\centering \textbf{Classic approach performance (runtime 1 hour)}
\end{tabular}
\begin{tabular}{p{35pt}p{35pt}p{29pt}p{37pt}p{25pt}p{30pt}} \hline
\centering &\textbf{\quad CPU usage \%} & \textbf{RAM usage \%} & \textbf{Memory usage \%} & \textbf{Utilized tweets} & \textbf{Accuracy \%} \\ \hline
\centering \textbf{Minimum} & 5.11 & 53.909 & 0.85 & \multirow{3}{*}{193,406} & \multirow{3}{*}{23.814} \\ \cline{1-4}
\centering \textbf{Average} & 6.18 & 54.109 & 0.922 &  & \\ \cline{1-4}
\centering \textbf{Maximum} & 13.76  & 54.716 & 30.683 &  & \\ \hline
\vspace{3cm}
\end{tabular}

\begin{tabular}{c}
\centering \textbf{Proposed approach performance (runtime 1 hour)}
\end{tabular}
\begin{tabular}{p{35pt}p{35pt}p{29pt}p{37pt}p{25pt}p{30pt}} \hline
\centering &\textbf{\quad CPU usage \%} & \textbf{RAM usage \%} & \textbf{Memory usage \%} & \textbf{Utilized tweets} & \textbf{Accuracy \%} \\ \hline
\centering \textbf{Minimum} & 4.92 & 42.007 & 0.144 & \multirow{3}{*}{167,322} & \multirow{3}{*}{36.273} \\ \cline{1-4}
\centering \textbf{Average} & 6.892 & 42.482 & 0.852 &  & \\ \cline{1-4}
\centering \textbf{Maximum} & 14.453  & 45.131 & 13.17 &  & \\ \hline
\end{tabular}\label{Exp_2}
\end{table}

\endgroup

\smallskip
\noindent{\bf \em $(i)$ First experiment result (Fixed running time).} 
In the first experiment, we check that there is almost no noticeable difference in CPU usage between the approaches as in Table \ref{Exp_2}. 
However, some comparable results are observed in RAM usage, where the classic approach's minimum usage is 53.9\%, average usage is 54.1\%, 54.7\% maximum, and the proposed approach's minimum usage level is 42\%, 42.4\% on average, and 45.1\% maximum, respectively. In the memory usage, we check that our approach is more efficient. 
Further, the most noticeable difference between the approaches is observed in the number of tweets utilized. During the 1-hour experimentation, the classic approach utilized 193,406 tweets, whereas the proposed approach utilized 167,322 tweets.  
\bnote{Although the number of tweets used in the proposed approach is approximately 26,000 smaller than in the classical approach,} it achieves a 36.2\% accuracy, whereas the accuracy of the classic approach is 23.8\%, which is 13\% less accurate than the proposed approach.

\smallskip
\noindent{\bf \em $(ii)$ Second experiment result (Fixed target accuracy).} 
Finally, we perform same experiment to achieve the target accuracy of Bitcoin price prediction. To do this, we set a target accuracy level of - VAF = 84.81\%, because we observed that the model with the proposed approach achieved this level of accuracy during the first experiment. We run both approaches until they reaches the target accuracy level and compare resource usage accordingly. 
We obtain our results in Figure \ref{Exp_3_Fig} and Table \ref{Exp_3_Tab}, respectively. 

\begingroup
\setlength{\tabcolsep}{4pt} % Default value: 6pt
\renewcommand{\arraystretch}{1.5} % Default value: 1
\begin{table}[ht]
\label{tab1}\centering 
\caption{Results from classic and proposed approaches after achieving the same target level of accuracy.}
%\vspace{0.2cm}
\begin{tabular}{c}
\centering \textbf{Classic approach performance (target accuracy is 84.81\%)}
\end{tabular}
\begin{tabular}{p{35pt}p{35pt}p{29pt}p{37pt}p{25pt}p{30pt}} \hline
\centering &\textbf{\quad CPU usage \%} & \textbf{RAM usage \%} & \textbf{Memory usage \%} & \textbf{Utilized tweets} & \textbf{Runtime} \\ \hline
\centering \textbf{Minimum} & 46.125 & 51.809 & 0.31 & \multirow{3}{*}{5,185,742} & \multirow{3}{*}{\thead{21h 36min\\39sec}} \\ \cline{1-4}
\centering \textbf{Average} & 61.832 & 53.387 & 9.48 &  & \\ \cline{1-4}
\centering \textbf{Maximum} & 85.671  & 56.129 & 18.53 &  & \\ \hline
\vspace{3cm}
\end{tabular}

\begin{tabular}{c}
\centering \textbf{Proposed approach performance (target accuracy is 84.81\%)}
\end{tabular}
\begin{tabular}{p{35pt}p{35pt}p{29pt}p{37pt}p{25pt}p{30pt}} \hline
\centering &\textbf{\quad CPU usage \%} & \textbf{RAM usage \%} & \textbf{Memory usage \%} & \textbf{Utilized tweets} & \textbf{Runtime} \\ \hline
\centering \textbf{Minimum} & 4.334 & 48.377 & 0.15 & \multirow{3}{*}{785,469} & \multirow{3}{*}{\thead{4h 17min\\14sec}} \\ \cline{1-4}
\centering \textbf{Average} & 7.758 & 51.624 & 5.712 &  & \\ \cline{1-4}
\centering \textbf{Maximum} & 16.631  & 53.865 & 17.728 &  & \\ \hline
\end{tabular}\label{Exp_3_Tab}
\end{table}

\endgroup

As a result, we first see that there is a significant difference in CPU usage in this experiment. 
In the classic approach, the CPU workload is between 46.1\% and 85.6\%, with an average of 61.8\%. The proposed approach shows a minimum of 4.3\%, average of 7.7\%, and maximum of 16.6\%, which is almost 9 times less than the classic approach used.
In the RAM usage, we check that there is no significant differences whereas we see that the average usage of memory in the proposed approach is better than that of classic one. 
Finally, we check that the classic approach runs for 21 hour 36 minute 39 second to achieve the target accuracy, which is almost five times more than the time required to achieve the same level by spending 4 hour 17 minute 14 second with the proposed model. 

From the experiment results, we conclude that the proposed approach has much advantages over the classic approach. Considering the poster's tweets with the highest number of followers can lead to accurate prediction and prevent the computer from wasting its resources.

\section{DISCUSSION}

\bnote{In this work, we have checked that it is better for prediction performance and resource efficiency to extract and use data suitable for price prediction than to use all data in Bitcoin price prediction through tweeter data. In particular, for this purpose, even if only the attribute data of the most follower among the data on Twitter was used, the results were much better than the classic approach using all data. Furthermore, the model contributes to the literature on tweet sentiment studies and price prediction using reinforcement learning and provides reliable advice for further in-depth analysis. 

However, there exist some limitations to the considered approach in this paper.  First, we only used Twitter posted data to analyze people's feelings, which may be biased since not all crypto-traders express their opinions on Twitter. We realize that Bitcoin values are affected by a variety of variables that cannot be captured only through Twitter sentiments. Tweets and other social media (e.g., Reddit and Facebook) may be used to extract feelings in the real world, such as through news and other sources including photos and videos from YouTube or TV channels. Second, we analyzed the price prediction of Bitcoin by considering only four attributes of Twitter. Additional comparison results can be obtained by considering other attributes such as tweet language and tweet poster’s location. In the case of tweet language, most of the data is expressed in one language (e.g., English), so it will not significantly affect the price prediction. However, it may be interesting to see how data according to the tweet poster's location affects the Bitcoin price and prediction performance. Third, the algorithm for the predictive model can be modified by extending it to deep reinforcement learning algorithms. This has the advantage of being able to express the Q-function used in Q-learning more accurately with the deep learning method, so it is expected to help improve prediction performance. Finally, considering other sources for sentiment data and other types of cryptocurrencies could also increase the accuracy of predictions. All of these things could be our further research. 

}

%In addition to the prediction results, we analyzed the usage of computer resources (CPU, RAM, and memory) while experimenting, and compared the results when we performed the same experiment with the classic approach. Running the proposed and classic approaches for 1 h provided the results, as shown in Table \ref{Exp_2}. The most comparable results were noticed in RAM usage, where the classic approach used 54.0\% RAM on average, and the proposed approach used 42.5\% RAM. Despite the fact that the proposed approach utilized approximately 26,000 fewer tweets within the experiment time, it achieved 13\% more accurate predictions than the classic approach.
%Finally, we set a target level of accuracy for both approaches to achieve in aiming to get comparable results. A significant difference was observed in CPU workloads. The classic approach used approximately 62\% of the CPU, whereas the proposed approach used almost nine times less, at just 6.92\% of the CPU. Another noticeable fact was seen in the number of utilized tweets that both the approaches used to achieve the target level of accuracy. To achieve 91.54\% accurate prediction, 785,469 tweets were sufficient for the proposed approach, which is 66\% less than the classic approach that utilized almost 5.2 million tweets. Finally, the runtime also had a significant difference. The proposed approach took approximately 260 min to hit the target level of accuracy, while the classic approach took approximately 1,240 min, which is 4.8 times more.

\section{Conclusion}
In this paper, we considered Bitcoin price prediction based on Q-learning using tweet data. 
We analyzed the manner in which Bitcoin-related information on Twitter affects the actual Bitcoin price by considering four main attributes: number of followers of the poster, number of comments on tweets, number of likes, and number of retweets.
We predicted the actual Bitcoin price using a Q-learning method, and obtained the most valuable attributes with three reward functions. We verified that tweets with the most user-related attributes had the greatest effect on the future Bitcoin price. Next, we compare our approach with a classic approach where all Bitcoin-related tweets without being attribute-filtering, are uses as input data for the model, by analyzing the CPU workloads, RAM usage, memory, time, and prediction accuracy. We conclude that the proposed approach has much advantages over the classic approach.

\section*{References}
\def\refname{}

\end{document}

%% file: mymath.tex
\newcommand{\real}{{\mathbb R}}

\newcommand{\separator}{
  \begin{center}
    \rule{\columnwidth}{0.3mm}
  \end{center}
}

\def\ie{{\it i.e.}}

\newtheorem{definition}{Definition}

\newcommand{\expect}[1]{\mathbb{E}[ #1 ]}

\newcommand{\beq}{\begin{eqnarray*}}
\newcommand{\eeq}{\end{eqnarray*}}
\newcommand{\beqn}{\begin{eqnarray}}
\newcommand{\eeqn}{\end{eqnarray}}
\newcommand{\bemn}{\begin{multiline}}
\newcommand{\eemn}{\end{multiline}}

\def\cA{{\cal A}}

\def\cS{{\cal S}}